\documentstyle[aps,epsfig]{revtex}

\begin{document}
\title{Multi -vortex states of a thin superconducting disk in a step-like external
magnetic field}
\author{M. V. Milo\v{s}evi\'{c}\cite{aut2}, S. V. Yampolskii\cite{perm} and F. M.
Peeters}
\address{Departement \ Natuurkunde, Universiteit Antwerpen (UIA), \\
Universiteitsplein 1, \ B-2610 Antwerpen, Belgium}
\date{\today}
\maketitle

\begin{abstract}
The vortex states in a thin mesoscopic disk are investigated within the
phenomenological Ginzburg-Landau theory in the presence of a step-like
external magnetic field with zero average which could model the field
resulting from a ferromagnetic disk or a current carrying loop. The regions
of existence of the multi-vortex state and the giant vortex state are found.
We analyzed the phase transitions between them and found regions in which
ring-shaped vortices contribute. Furthermore, we found a vortex state
consisting of a central giant vortex surrounded by a collection of
anti-vortices that are located in a ring around this giant vortex.
\end{abstract}

\section{INTRODUCTION}

Recent progress in microfabrication and measurement techniques makes it
possible to study the properties of mesoscopic superconducting samples, and,
in particular, disks with sizes comparable to the penetration depth $\lambda 
$ and the coherence length $\xi $~\cite{Geim,Schw1,Schw2,Palas,Schw3,Sergey}%
. Depending on the disk size various superconducting states can exist in
such samples in an external magnetic field.

In small disks, where the confinement dominates, there exist only giant
vortex ($GV$) states (circular symmetric states with a fixed value of
angular momentum $L$)~\cite{Schw1,fink}. For sufficiently large disks the $%
GV $ state can break up into the multi-vortex ($MV$) state (a mixture of
giant vortices with different angular momentum)~\cite{Schw2,Palas}. Also
ring-shaped vortex states with larger energy than $GV$ and $MV$ ones were
predicted~\cite{ring}. These ring-shaped two-dimensional vortex states have
a cylindrically symmetric magnetic field profile and they are different from
the ring-vortices that were e.g. found in three-dimensional superfluid
liquid helium. But the analysis of Ref.~\cite{Sergey} showed that in a
superconducting disk in the presence of a homogeneous magnetic field the
ring-shaped vortices are unstable.

\begin{figure}[tbp]
\begin{center}
\epsfig{file=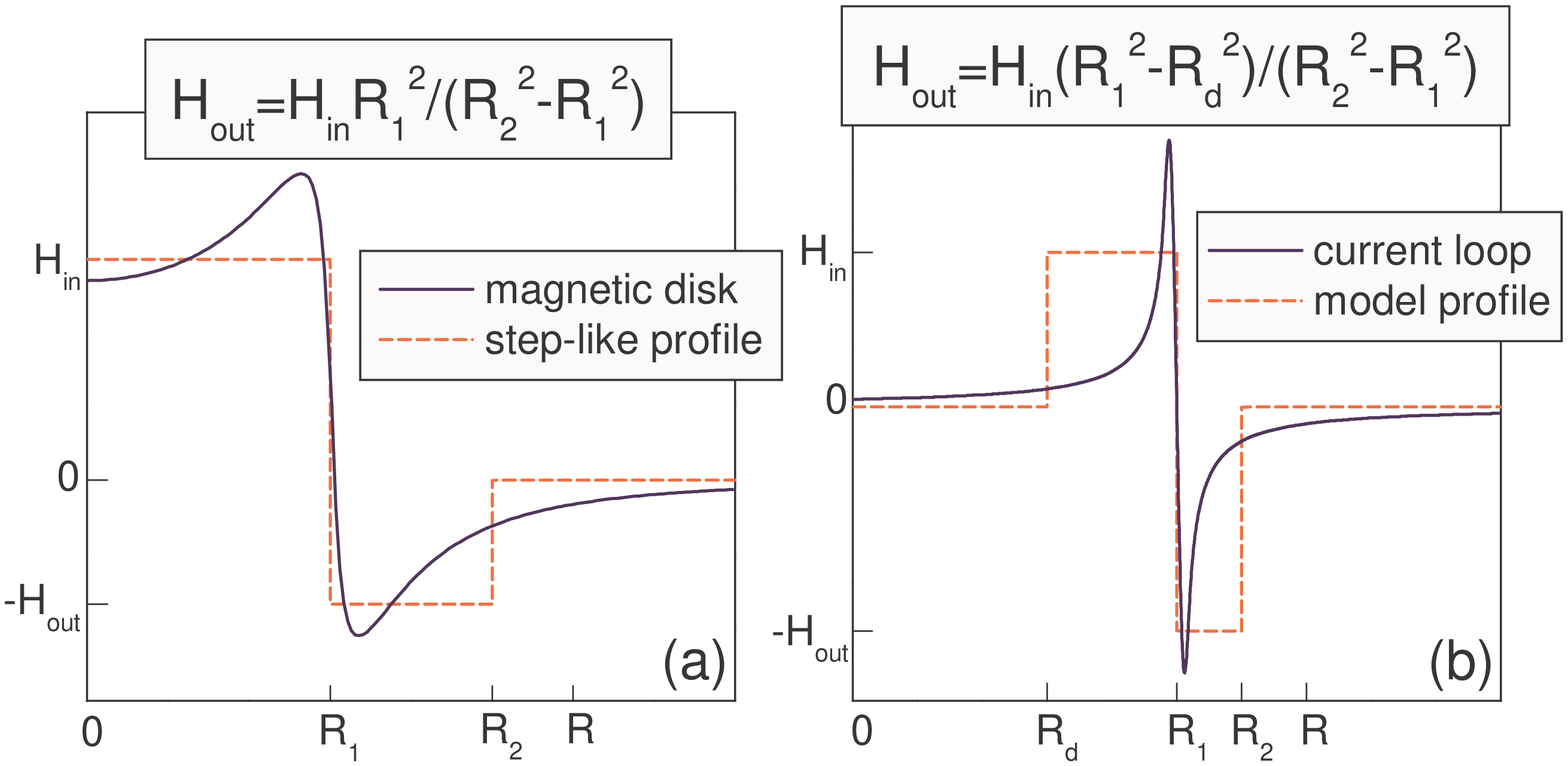, width=120mm, height=90mm, clip=}
\end{center}
\vspace{-3cm}
\caption{The magnetic field profile from a magnetic disk (a) and a current
loop (b) and the corresponding step field models.}
\label{fig1}
\end{figure}

In the present paper we study the $MV$ states of a thin superconducting disk
in an inhomogeneous magnetic field $\vec{H_{0}}$ with step-like
profile. This magnetic field profile could approximate the field resulting
from a magnetic dot (they are usually used in experiments, e.g. see Ref.~\cite{leuven}) as well as the field of a current loop (Fig.~\ref{fig1}). We
found that this model field profile captures a lot of interesting physics
and that the step height and width strongly influence the superconducting
phase diagram.

\vspace{-0.8cm}

\section{APPROACH USED}

The details of our approach can be found in Refs.~\cite{Schw2,Palas,Sergey}
and, therefore, we briefly sketch only the necessary steps. We restrict
ourselves to sufficiently thin disks such that $d<<\xi ,\lambda $. In this
case, to a first approximation, the magnetic field due to the circulating
superconducting currents may be neglected and the magnetic field inside the
disk equals the external one. Following Refs.~\cite{Schw2,Palas,Schw3,Sergey}%
, the order parameter of the $MV$ state is written as a linear combination
of eigenfunctions of the {\it linearized} Ginzburg-Landau equation 
\begin{equation}
\psi \left( \vec{\rho }\right)
=\sum_{L_{j}=0}^{L}\sum_{n=0}^{\infty }C_{n,L_{j}}f_{n,L_{j}}\left( \rho
\right) \exp \left( iL_{j}\varphi \right) \text{,}  \label{lin}
\end{equation}
where $\rho $ is the radial distance from the disk center, $\varphi $ is the
azimuthal angle, and $n$ enumerates the different radial states for the same 
$L_{j}$. In the homogeneous magnetic field case $L$ is the value of the
effective total angular momentum which is equal to the number of vortices in
the disk. Later we will see that, for our inhomogeneous magnetic field case,
the assignment of the total vorticity can be tricky.

Substituting Eq.~(\ref{lin}) in the free energy (measured in $%
F_{0}=H_{c}^{2}V/8\pi $ units) 
\begin{equation}
F=\frac{2}{V}\left\{ \int dV\left[ -\left| \psi \right| ^{2}+\frac{1}{2}%
\left| \psi \right| ^{4}+\left| -i\vec{\nabla }\psi -%
\vec{A}\psi \right| ^{2}+\kappa ^{2}\left( \vec{h}%
\left( \vec{\rho }\right) -\vec{H_{0}}\right) ^{2}%
\right] \right\} \text{,}  \label{free_en}
\end{equation}
(here $\vec{h}\left( \vec{\rho }\right) =%
\vec{\nabla }\times \vec{A}\left( \vec{\rho 
}\right) $ is the local magnetic field, $\vec{A}$ is the vector
potential) we obtain $F$ as a function of the complex parameters $%
C_{n,L_{j}} $. Minimization of $F$ with respect to these parameters allows
to find the equilibrium vortex configurations and to determine their
stability.

We restrict ourselves to states built up by only two components in Eq.~(\ref
{lin}). It brings quantitative bounds in our analysis but, nevertheless,
gives the correct qualitative behavior and facilitates the physical insight
into the problem. Using two different approaches from Refs.~\cite
{Schw3,Sergey} $MV$ configurations with three and five components in Eq.~(\ref
{lin}) were checked. This resulted in 1)~a reduction of the region of
existence of metastable $MV$ states at the low magnetic field limit, 2)~an
appearance of additional unstable states corresponding to the saddle points
of the free energy function, and 3)~no new vortex configurations in the
ground state. Thus, two components in Eq.~(\ref{lin}) are sufficient to
describe the vortex structure in the ground state of our system. Depending
on the radial state of the components we defined two different
configurations: {\it giant-giant} ($n_{1}=n_{2}=0$) and {\it giant-ring} ($%
n_{1}=0,$ $n_{2}=1$) $MV$ states. The first represents a combination of two $%
GV$ states while the second is composed of one giant and one ring vortex.

\section{CENTERED STEP-LIKE MAGNETIC FIELD}

\begin{figure}[tbp]
\begin{center}
\epsfig{file=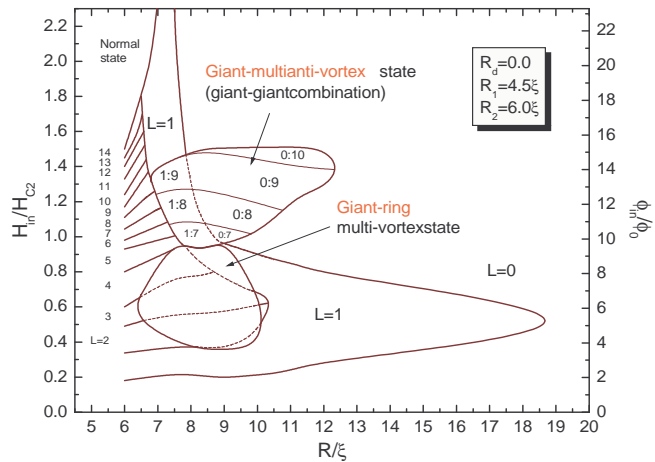, width=124mm, height=93mm, clip=}
\end{center}
\vspace{-3cm}
\caption{The $H_{in}-R$ phase diagram for the ground state of a thin
superconducting disk in a step-like magnetic field. Solid curves indicate
transitions between different vortex states including different $MV$
regions. Dashed curves denote the transitions between different metastable $%
GV$ states.}
\label{fig2}
\end{figure}

First, we investigate the case of a centered step-like magnetic field
corresponding to a magnetic disk. The effect of the size of the
superconducting disk on the vortex state is shown in the phase diagram of
Fig.~\ref{fig2}. The parameters considered are $R_{1}=4.5\xi $ and $%
R_{2}=6.0\xi $. The solid lines indicate where the ground state of the free
energy changes from one state to another and dashed lines correspond to
transitions between different $GV$ states as metastable states. The right
axis of the diagram shows the corresponding positive flux through the region 
$\rho <R_{1}$.

\vspace{-0.5cm}

One can clearly see the {\it re-entrant} behavior. For example, for $%
R=9.0\xi $ we observe the following change of total vorticity $%
L=0\rightarrow 1\rightarrow 2\rightarrow 3\rightarrow 1\rightarrow 0$.
Notice that the ground state with total vorticity $L=0$ and $L=1$ covers the
largest part of the phase diagram. With increasing disk size, all other $GV$
states are strongly suppressed in favor of the various $MV$ states. For
small disk sizes different $GV$ states are present as the ground state. But
with increase of disk size, islands with different $MV$ configurations ({\it %
giant-ring} and {\it giant-giant} $MV$ states) dominate the ground state
diagram. For $R=6.38\xi $, a giant-ring $MV$ state appears, and for $%
R=6.7\xi $ we obtain giant-giant states.

\begin{figure}[tbp]
\begin{center}
\epsfig{file=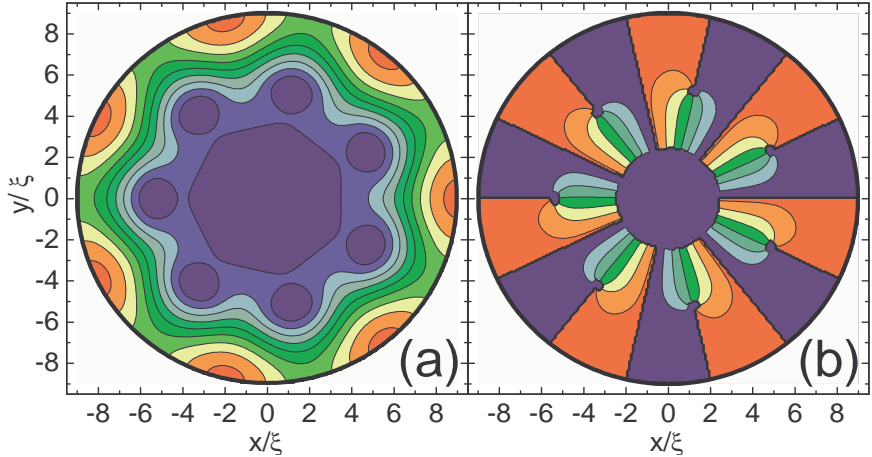, width=120mm, height=90mm, clip=}
\end{center}
\vspace{-2.5cm}
\caption{(a) Contour plot of the superconducting density for the $(0:7)$ 
{\it giant--multi anti-vortex} ground state, and (b) the corresponding
contour plot of the phase. Notice that the phase near the boundary is near $%
0 $ (blue) or $2\protect\pi $ (red) but due to the finite numerical accuracy it
oscillates between $2\protect\pi -\protect\varepsilon $ and $2\protect\pi +%
\protect\varepsilon $ where $\protect\varepsilon \approx 10^{-5}$.}
\label{fig3}
\end{figure}

As one can see in Fig.~\ref{fig2}, for example, for $R=9.0\xi $ and with
increasing magnetic field $H_{in}$, giant-giant combinations $%
(L_{1}:L_{2})=(0:7)$, $(0:8)$, $(0:9)$ etc. become the ground state. In the
contour plot of the $\left| \psi \right| ^{2}$-distribution for these states
we observe single vortices which are arranged on a ring with a low density
area (see Fig.~\ref{fig3}(a), blue color - zero density, red - highest density) situated in the center of the disk (see Fig. \ref{fig3}(a), for the $%
(0:7)$ state). This central area is associated with a giant vortex, and
encircling it leads to a phase change showing the vorticity $L=7$.

Fig.~\ref{fig3}(b) shows the contour plot of the corresponding phase with
seven anti-vortices arranged in a circle around the giant vortex and the 
{\it total} vorticity is $L=0$. It is important to emphasize that for the
disk size $R=R_{2}$ and with changing of $R_{1}$ {\it giant-giant} $MV$
states consist only of vortices arranged radially symmetric with no
anti-vortices present. Thus, it is obvious that enlarging the
superconducting disk enhances the influence of the negative part of the
step-like magnetic field. As a result we obtain a new giant-giant form
called {\it giant--multi anti-vortex} states. The total vorticity is now
equal to the lowest vorticity of the giant vortex states of which the
multi-vortex consists of. This inevitably leads to a re-entrant behavior in
the total vorticity, which is in agreement with Ref.~\cite{Schw4} where it
was found numerically that several vortices can enter (or exit) at once for
disks with sufficiently large radius. For $R=R_{2}$ disks we only found that 
$L\rightarrow L\pm 1$ transitions are possible. With increase of the disk
radius the $(0:8)$ and $(0:9)$ {\it giant-multi anti-vortex} states become $%
(1:8)$ and $(1:9)$, respectively. For $R>12.33\xi $, the multi-vortex states
become metastable again and the $L=0$ and $L=1$ states are now the only
ground states. Moreover, for $R>18.67\xi $ we have the Meissner state for
all values of the applied magnetic field.

\section{SHIFTED STEP-LIKE MAGNETIC FIELD}

\begin{figure}[tbp]
\begin{center}
\epsfig{file=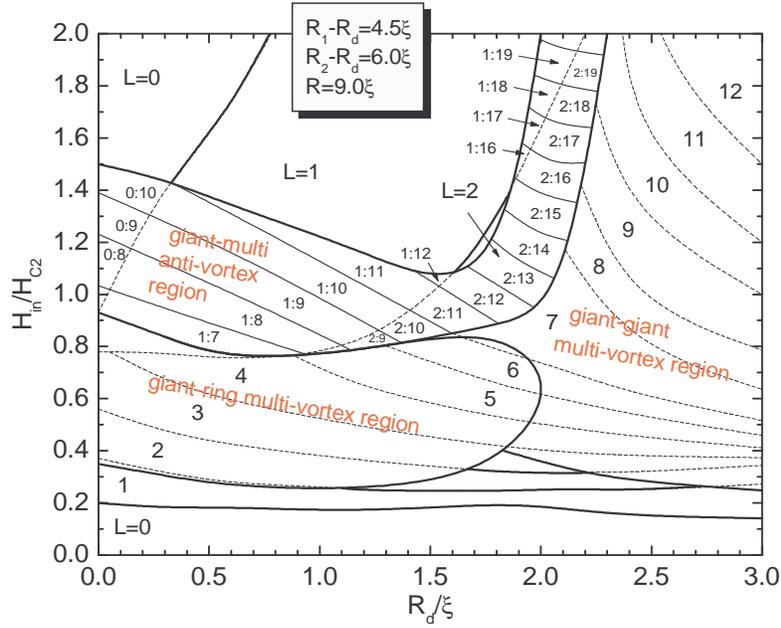, width=160mm, height=120mm, clip=}
\end{center}
\vspace{-3.5cm}
\caption{The $H_{in}-R_{d}$ phase diagram for the ground state of a thin
superconducting disk in a ring step-like magnetic field distribution. The
same curve convention is used as in Fig.~\ref{fig2}.}
\label{fig4}
\end{figure}

Next, we move the step-like field profile along the radius of the
superconducting disk, i.e. consider a ring magnetic field profile as in the
case of a current loop. In Fig.~\ref{fig4}, we present the phase diagram as
function of $R_{d}$. The parameters of the magnetic field profile are $%
R_{1}-R_{d}=4.5\xi $, $R_{2}-R_{d}=6.0\xi $, and $R=9.0\xi $. Thick solid
curves indicate the transitions between different vortex states and dashed
lines denote transitions between metastable $GV$ states.

\begin{figure}[tbp]
\begin{center}
\epsfig{file=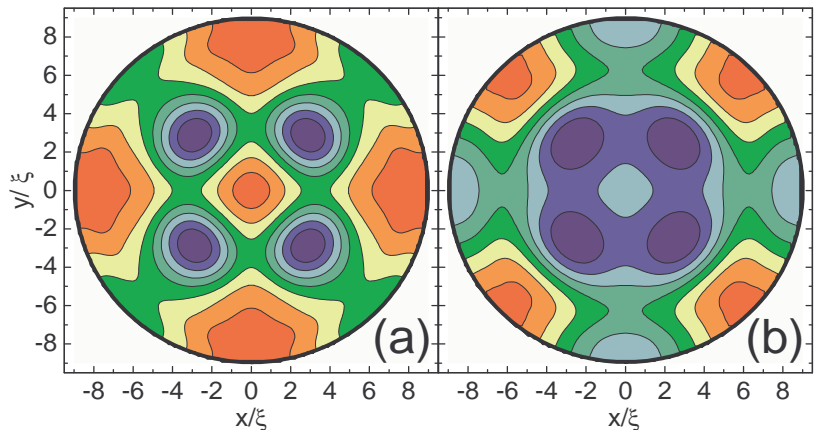, width=112mm, height=84mm, clip=}
\end{center}
\vspace{-2.2cm}
\caption{Contour plots of the superconducting wave density for {\it %
giant-giant} (a) and {\it giant-ring} (b) $MV$ state with the total
vorticity $L=4$.}
\label{fig5}
\end{figure}

The {\it re-entrant} behavior, as in previous case, is clearly visible. The
most important result is that the $MV$ states dominate this diagram.
Moreover, we have both types of multi-vortices, i.e. {\it giant-giant} and 
{\it giant-ring}, as ground state, and with shifting the field profile
towards the disk periphery, giant-giant $MV$ configurations cover most of
the superconducting region. The giant-giant $MV$ states in this case appear
both as states with no anti-vortices present, and as {\it giant--multi
anti-vortex} states. Following the transition lines, a correspondence
between different giant-multi anti-vortex states $(L_{1}:L_{2})$ can be
seen: $L_{2}$ remains the same, while $L_{1}$ increases from $0$ to $2$. As
one can see, the latter is strongly correlated with {\it re-entrant}
behavior. In the rest of the diagram, the $MV$ states with the ''classical''
geometry (both {\it giant-giant} and {\it giant-ring}) dominate. However, a
difference between those states exists. As shown in Fig.~\ref{fig5}, in
the case of a {\it giant-ring} state, vortices are preferentially
distributed within a ring-shaped lower density area. Further, considering
the high density areas at the disk periphery, a shift in phase of $\Delta
\theta =\pi /L$ is observed, where $L$ is the total vorticity. Although
these states have a different origin sometimes they can exhibit a similar
distribution of vortices but the phase is always able to distinguish between
them.

\vspace{-0.2cm}

\bigskip

\begin{center}
{\large ACKNOWLEDGEMENTS}

\bigskip
\end{center}

This work was supported by the Flemish Science Foundation (FWO-VI), the
Belgian Inter-University Attraction Poles (IUAP-IV), the ``Onderzoeksraad
van de Universiteit Antwerpen'' (GOA), and the ESF programme on ``Vortex
matter''.

\end{document}